# Single nitrogen vacancy centers in chemical vapor deposited diamond nanocrystals

J. R. Rabeau[a]

*Department of Physics, Division of Information and Communication Science, Macquarie University, New South Wales 2109, Australia.*

A. Stacey, A. Rabeau, S. Prawer

*School of Physics, Microanalytical Research Centre, the University of Melbourne, Parkville, Victoria 3010, Australia.*

F. Jelezko, I. Mirza, J. Wrachtrup

*3. Physikalisches Institut, Universität Stuttgart, Pfaffenwaldring 57, 70569 Stuttgart, Germany.*

a) corresponding author: jrabeau@ics.mq.edu.au

Abstract

Nanodiamond crystals containing single color centers have been grown by chemical vapor deposition (CVD). The fluorescence from individual crystallites was directly correlated with crystallite size using a combined atomic force and scanning confocal fluorescence microscope. Under the conditions employed, the optimal size for *single* optically active nitrogen-vacancy (NV) center incorporation was measured to be 60 to 70 nm. The findings highlight a strong dependence of NV incorporation on crystal size, particularly with crystals less than 50 nm in size.

Optically active defects in diamond offer many potential uses in quantum information processing. For example, the nitrogen-vacancy (NV) color center in diamond has been demonstrated to be an efficient source for single photons[1,2] and has been used to implement quantum key distribution in free space.[3] It is also being explored as a spin-based qubit in quantum computing.[4] The nickel-related color centers[5,6,7] and silicon-vacancy center (SiV)[8] also show increasing promise as optical qubits for their narrow line-widths. More recently, a myriad of new experiments employing nanodiamonds as fluorescence labels (using multiple color centers instead of single centers) in biological systems have emerged.[9,10] Key advantages of nanodiamonds as compared to quantum-dots or other conventional fluorescence bio-labels are the non-cytotoxicity[11], room temperature photostability[12] and relative ease with which surfaces can be functionalized.[13] Although the bio and quantum-related areas are fundamentally different, the underpinning materials requirements have significant overlap. Quantum applications require spectrally and temporally stable emission from single defects, and biological applications demand bright fluorescence from small crystals, for example. Although the end-uses are different, the current limitations in both fields consistently point back to the fundamental materials science questions. For example, open issues include: the crystal size dependence of fluorescence from defect centers, the effect of surface termination as crystals become smaller and the mechanisms of color center formation during nanodiamond growth.

Incorporation of nitrogen in the diamond lattice is energetically favorable, and therefore nearly all natural and synthetic diamond contains some native concentration of N. Consequently, N is in some way involved in the majority of the defect centers studied in bulk diamond.[14] The majority of single defect experimentation and implementation has been limited to natural and high-pressure high-temperature (HPHT) single crystal "bulk" diamond or detonation produced nanodiamond powder. Drawbacks related to these materials stem from the relatively high background fluorescence of some bulk materials and, except for the case of ion implantation, limited control over defect incorporation. Electron irradiated nanodiamond powder, ground to an average size of 50 nm,[15] has been shown to contain single NV centers, but the powder form and propensity for agglomeration makes this material challenging to work with. For these reasons, it is important that a more robust and versatile technique for single color center incorporation in diamond be developed. Chemical vapor deposition (CVD) is a promising alternative because diamond can be grown on numerous substrate materials, dopants and crystal size can be carefully controlled, and parallel fabrication strategies can be implemented. However, an important result not yet reported using the CVD technique is the controlled synthesis of single optically active NV centers. This capability may be critical in the search for a repeatable and scalable fabrication strategy for quantum information proposals which employ NV centers. With respect to materials processing, this has proven to be challenging for a number of reasons: firstly, because nitrogen is so abundant in the atmosphere, it requires great care to remove all unwanted residual nitrogen from a typical diamond growth chamber. Therefore, during growth, it is difficult to inhibit the incorporation of NV centers, which form spontaneously during CVD diamond synthesis, to a level where single defects cannot be individually resolved. Second, the presence of non-diamond ($sp^2$) carbon and other defects in CVD diamond gives rise to strong, spectrally broad background fluorescence. As a result, the background fluorescence often swamps any would-be single photon emission despite careful optical filtering. We solved both of these problems by growing isolated diamond nanocrystals which minimizes the volume of diamond available for color center incorporation, and eliminates a large proportion of unwanted background fluorescence.

Diamond nanocrystals were grown on quartz cover slips using a 1.2 kW microwave plasma chemical vapor deposition reactor (ASTeX). The chamber pressure was maintained at 40 mbar with a 0.7% $CH_4$ in $H_2$ gas mixture. The substrate temperature was 800ºC during the growth period. For the growths reported here, nitrogen was not deliberately added to the gas feedstock as it is known to be present at a background level of ~0.1%, which corresponds to an N/C ratio of 0.15.

Prior to growth, the substrates were seeded by exposure to a suspension consisting of metal and diamond powder (<250 nm, De Beers) in an ultrasonic bath. For the purpose of these experiments, a relatively low nucleation density was required, combined with a very short growth time. The typical sample consisted of nanodiamond particles on the quartz substrate separated by 1 – 20 um.

The samples were characterized with a lab-built, room-temperature confocal sample-scanning fluorescence microscope (100X oil immersion objective lens, 1.4 NA) combined with a commercial AFM system. Excitation of the NV center was achieved with a 532 nm CW diode pumped solid state laser (Coherent, Compass). For detection of red-shifted fluorescence the pump beam was blocked with an interference long-pass filter to select only light with wavelength above 650 nm for detection. A low-temperature beam-scanning confocal microscope operating with the sample immersed in super fluid helium was used for characterization of fluorescence excitation lines of NV centers at cryogenic temperature.

The combined AFM/confocal system was aligned such that the laser spot coincided with the AFM cantilever tip. This enabled unambiguous identification of those crystals which gave rise to NV fluorescence. Crystals which did fluorescence were further analyzed to establish if one or more NV centers were present in the crystal. The latter measurement was accomplished by directing the fluorescence into a Hanbury Brown and Twiss interferometer to measure the second order correlation function ($g^{(2)}(\tau)=\langle I(t)I(t+\tau)\rangle/\langle I(t)^2\rangle$). For a delay time $\tau = 0$, $g^{(2)}$ shows an "antibunching" dip indicating sub-Poissonian statistics of the emitted photons. This occurs because a single quantum system cannot simultaneously emit two photons. The contrast in $g^{(2)}$ scales as 1/N, where N is the number of emitters. Note that the deviation from maximum contrast for a single defect results from the background fluorescence from the quartz substrate. This minor background contribution was constant through all measurements and originated primarily from the substrate.

Figure 1 shows a typical dataset where (a) is the AFM height data and (b) is the corresponding confocal fluorescence data. Bright spots in the confocal image have corresponding features in the AFM image which enables precise analysis of emission as a function of crystal size. Figure 2 is a compilation of second order correlation histograms from 3 different emitters in the sample. Curve C clearly shows $g^{(2)} < 0.5$ at $t = 0$ indicating a single NV center. Curves A and B correspond to crystals containing more than 1 emitter.

Analysis of ~20 crystals showing NV fluorescence gave an indication as to the distribution of single and multiple NV's and their relationship to crystal size. Figure 3 shows a histogram of height distribution in this diamond sample. Scanning electron microscopy of these samples revealed an approximately spherical shape and therefore the "height" from the AFM measurements refers to the crystal diameter. The feature sizes ranged up to 120 nm. Superimposed over this histogram of *all* crystals measured (even those without any fluorescence) is the histogram of the probability for 1, or more than 1 NV centers to be present in a given crystal size range. As is shown, the range in which *single* NV's are preferentially incorporated is between 60 and 70 nanometers. As the crystal size increases, so too does the number of emitters per crystal.

We now discuss the results in light of the observation that, under the conditions employed here, nanocrystals smaller than about 40nm do not have any observable NV emission.

The incorporation fraction, F, of nitrogen in diamond, is given by:

$$F = \frac{[N]_{gas}/[C]_{gas}}{[N]_{film}/[C]_{film}}$$

where $[N]_{gas}/[C]_{gas}$ is the atomic ratio of nitrogen to carbon in the gas phase and $[N]_{film}/[C]_{film}$ is the ratio of nitrogen to carbon in the deposited film. The latter has been measured for neutral substitutional N by a number of authors and has been found to vary from $1.6 \times 10^{-4}$ for homoepitaxial growth[16] to $5 \times 10^{-5}$ for CVD growth under conditions similar to those reported here[17]. In the present work, the ratio $[N]_{gas}/[C]_{gas} = 0.15$ and therefore the concentration of substitutional N in the nanoparticles is expected to be of the order of 15 ppm. The ratio of NV to substitutional nitrogen is not known precisely, however, based on ESR and optical measurements Vlasov *et al.* [18] have estimated this ratio to be less than $10^{-3}$. Therefore the concentration of active NV is expected to be of the order of 0.015ppm in our nanocrystals.

Based on the above reasoning the inset to Fig 3 shows the expected number of NV centers per nanocrystal as a function of nanocrystal diameter. It can be seen that on average a 60 to 70nm diameter crystal would be expected to have 0.15 to 0.2 NV centers, i.e. the probability of observing an NV in such a crystal is about 15-20%; this should be compared to the 2% value deduced from our measurements. The fact that the probability of observing a single NV in a nanocrystal is lower than what we expect may indicate that other factors are at work, such as for example the proximity of the NV to surface defects may play a role in quenching NV emission. Furthermore, it has been shown by density functional tight binding simulations that nitrogen preferentially resides near the surface of nanodiamond particles and not in the core.[19] This would lead to a slight overestimate of the concentration of N within our nanodiamonds, and therefore we should actually expect fewer NV centers.

It can be seen that the probability of observing an NV decreases very rapidly as the crystal size is reduced, which agrees with the measured low probability to observe NV emission from diamond nanocrystals less than 40nm or so in diameter. On the other hand the probability of NV emission increases very sharply as the crystal size grows and for 110nm diameter crystals, the probability of observing at least one NV center per crystal approaches unity. Although the absolute numbers quoted above may be subject to some degree of error, our data do appear to be consistent with the proposition that (i) the conversion of N from the gas phase into NV is very low (of the order of $5 \times 10^{-8}$) and (ii) the probability for observation of an NV center in a diamond crystal depends primarily on the number of N atoms within the crystal.

It would be tempting to assume that the number of NV's could be increased by simply increasing the N/C ratio in the gas phase. However, when the [N]/[C] ratio exceeds a concentration of about 0.2, the increase of nitrogen in the gas phase degrades the quality of the CVD nanocrystals without any substantial increase in the concentration of N in the film. Nevertheless, in order to more deeply understand N and NV incorporation, we have undertaken to perform similar experiments as a function of N-addition during growth.

The single color centres grown in diamond nanocrystals were now studied in more detail in terms of spectral emission properties and spin coherence lifetime, both of which are relevant in quantum information science. At cryogenic temperatures optical transitions of single NV defects in high quality bulk crystals are shown to be narrower compared to splitting between the spin sublevels.[20] Hence the frequency of the emitted photons carries information about the spin state. This relationship between spin state and properties of the emitted photon provide the possibility for distant entanglement generation between NV centers, with potential applications in quantum communications.[21,22]

As usual for solid state systems, optical properties of the dopants are strongly dependant on the quality of the host crystals. It has been demonstrated that nitrogen-free diamond crystals show a nearly transform-limited linewidth of the optical transitions.[23] The presence of nitrogen in the host usually leads to broadening of spectral lines. This effect was attributed to ionization of nitrogen under 1.945 eV illumination. Such ionization creates a fluctuating electric field at the location of th NV center causing spectral diffusion[20] and thus broadening.

Low temperature properties of optical transitions were examined using fluorescence excitation spectroscopy. In our experiments the sample was cooled down to T= 1.6 K and a narrow band laser (linewidth <1 MHz) was swept across the zero-phonon line of the NV defect. The fluorescence emitted into the phonon sideband was detected. Figure 4 shows a typical example of fluorescence excitation which is the result of averaging over 40 laser sweeps. The spectral line, which is 11 GHz, shows significant broadening compared to the lifetime-limited value (12 MHz). This broadening is possibly related to ionization of other than NV impurities in the diamond crystal causing spectral jumps. Such fluctuations, when averaged, result in a broad Gaussian lineshape (see Figure 4). Note that the excitation spectrum was recorded using a resonant laser only (637 nm wavelength). Typically, bulk crystals show irreversible fading at low temperature when excited at 637 nm and application of a continuous repumping green or blue laser beam is necessary to bring the center back into the fluorescing state.[24] Possibly this process is related to photoionization of the NV defect via a metastable state. The repuming green beam ionizes the electron acceptors which lead to recharging of the NV center back to the negative charge state. The absence of irreversible photoionization in nanocrystals demonstrates that the acceptors responsible for fading of the NV defect can be repumped with a red laser resonant with the NV defect. Hence, NV photostability in CVD grown nanocrystals does not require multicolor repumping as is the case with bulk diamond.

Finally, the spin coherence lifetime ($T_2$) of the NV center is interesting in the context of spin based quantum coherence applications. In order to probe the $T_2$ time, the spin state evolution under a Hahn echo pulse sequence has to be investigated. The classical Hahn echo pulse sequence consists of $\pi$ and $\pi/2$ microwave pulses and a waiting time $\tau$. For optical detection, an additional $\pi/2$ pulse was introduced in the echo sequence to convert the spin echo phenomenon into optically measurable populations. $T_2$ has been measured previously to be several μs for NV centers in nanodiamond powders[25] and in single crystal diamond the longest recorded $T_2$ time recorded has been 350 μs.[26] For the CVD nanocrystals studied here, the Hahn echo decay from a single NV center is shown in Figure 5. $T_2$ was measured to be up to 15 μs which is in agreement with what was previously measured in nanodiamond powders. In comparison to 350 μs in bulk diamond, there is clearly some decoherence mechanism at work which causes the large change in $T_2$. Nitrogen, which is a spin ½ system, is generally accepted to be a primary source of decoherence, however the $^{13}C$ isotope, in it's natural abundance (1%) also contributes and is in fact the limiting factor in type IIa single crystal diamonds. By reducing the base N concentration, it is possible that the coherence lifetime can be increased. Superimposed on the echo decay which marks $T_2$, a strong modulation can be observed. This echo envelope modulation is a well known phenomenon from ESR measurements on NV defects. It results from a coupling of the NV electron spin to the neighboring $^{13}C$ spins.[27]

In this work we have demonstrated, for the first time, CVD growth of isolated diamond nanocrystals containing single optically active NV defects confirmed by antibunching experiments. The nanodiamonds were characterized to establish the size dependence and probability of optically active NV incorporation and it was found that under the conditions employed, single optically active NV centers preferentially incorporate in crystals between 60 and 70 nm and that the proportion of N-atoms in the gas phase which form NV centers is ~$5 \times 10^{-8}$.

The spectral stability was measured using resonant photoluminescence excitation at 1.6 K. As anticipated, the fluorescence from a single NV centre was spectrally broad (~11 GHz), however it was observed that re-pumping of the NV centre with a weak green laser was not required.

The ground state spin coherence lifetime was measured to be 15 μs which is short compared with ultrapure single crystal diamond where the coherence time reaches the limit imposed by intrinsic $^{13}C$ nuclear spins. This may in part be due to the high concentration of substitutional N in the diamond.[28] The coherence lifetime may be improved by reducing the concentration of N and $^{13}C$ in the CVD process gases.

The capability to grow isolated nanocrystals containing single color centers on different substrates opens up a broad range of applications in the field of quantum information science. Understanding fundamental materials science issues such as size dependence, dopant requirements and decoherence mechanisms are vital for continued success in this area. Indeed there is strong incentive to pursue this with respect to spin based quantum coherence applications using nanodiamond instead of bulk single crystal diamond. Furthermore, the use of nanodiamonds in fluorescence applications for biological and life sciences shows great promise. Analysis tools such as those discussed in this report are now essential for progress in both fields of research.

Acknowledgements

We thank Alberto Cimmino for useful discussion on AFM analysis and Jim Butler for comments on diamond materials science. This work was funded by the Australian Research Council Discovery Projects scheme, the Department of Education Science and Training International Science Linkages program established under the Australian Government's innovation statement Backing Australia's Ability, and the EU (QAP, EQUIND, NEDQIT); DFG (SFB/TR 21), and Landesstiftung BW Foundation (Atomoptik).

[27] L. Childress, M. V. Gurudev Dutt, J. M. Taylor, A. S. Zibrov, F. Jelezko, J. Wrachtrup, P. R. Hemmer, and M. D. Lukin, Science, DOI: 10.1126/science.113187 (2006).

[28] J. Wrachtrup and F. Jelezko, J. Phys. Cond. Mat. 18, S807 (2006).

Figure captions

FIG. 1. a) Atomic Force Microscopy image of nanocrystalline diamond deposited on quartz substrates. b) The corresponding confocal fluorescence image of the same region of the sample. The bright fluorescing spots indicate emission from NV centers in the diamond.

FIG 2. Antibunching from 3 different crystals in the sample, showing emission from single (C) and multiple NV centers.

FIG. 3. Histogram of the overall sizes of the CVD diamond nanocrystals (primary Y-axis) and a histogram of the probability for a given crystal size range to contain 1 or more than 1 NV color centers (secondary Y-axis). This information was measured using a Hanbury-Brown and Twiss interferometer. The inset shows the expected number of NV centers within a given crystal size according to literature values for N-incorporation efficiency in CVD diamond, and the conversion efficiency of N to NV.

FIG 4. Fluorescence excitation of a single NV defect in a CVD diamond nanocrystal at T= 1.6 K. The red solid line is the Gaussian fit indicating a linewidth of 11 GHz.

FIG. 5. Hahn echo decay curve of a single NV center in a diamond nanocrystal. The echo amplitude was measured as a function of delay between ESR pulses. The spin coherence lifetime was measured to be ~10 μs. The modulation superimposed on the echo decay curve results from the coupling of the electron spin of NV defect to 13C nuclear spins.

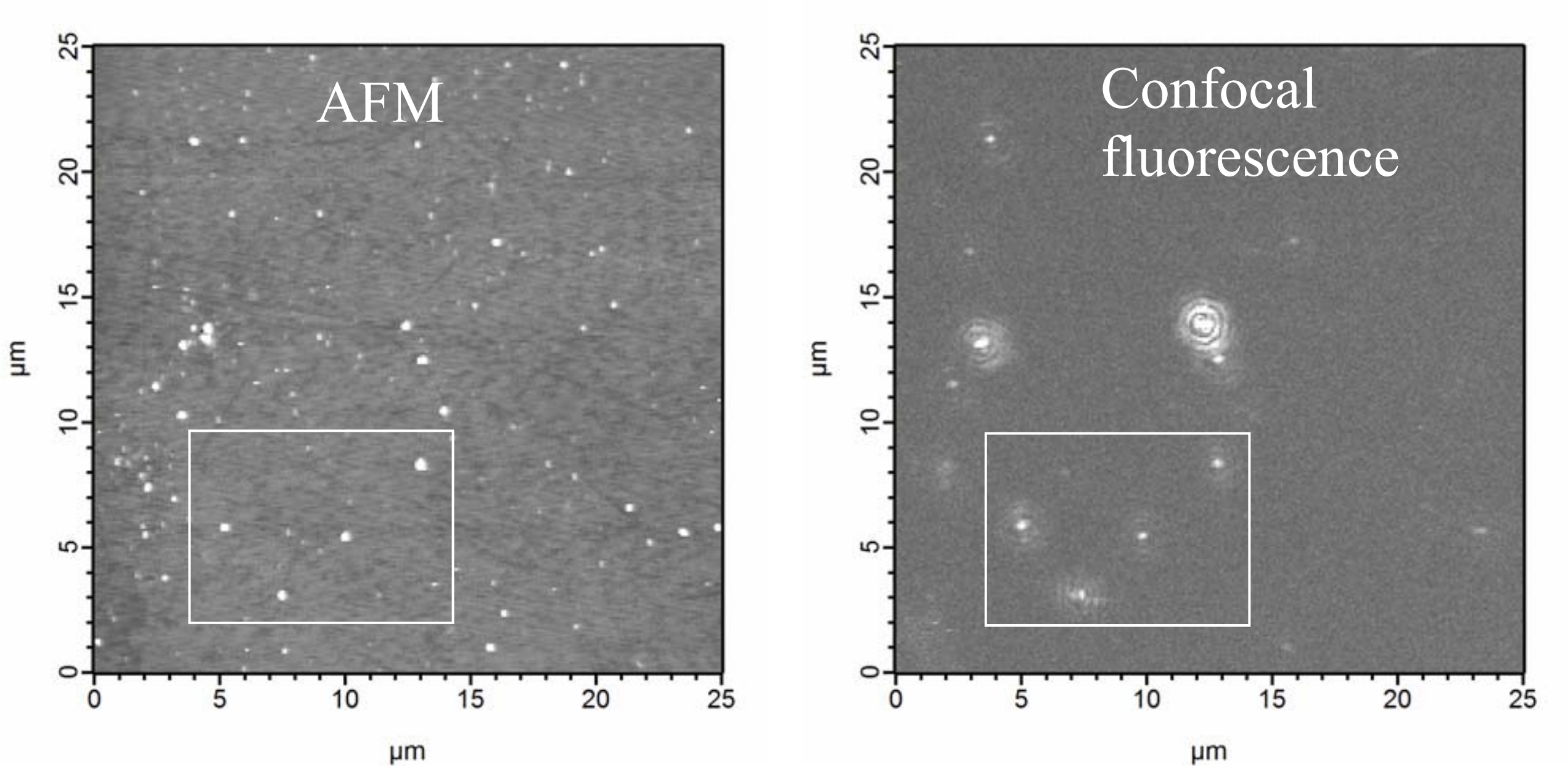

FIG 1. J. R. Rabeau

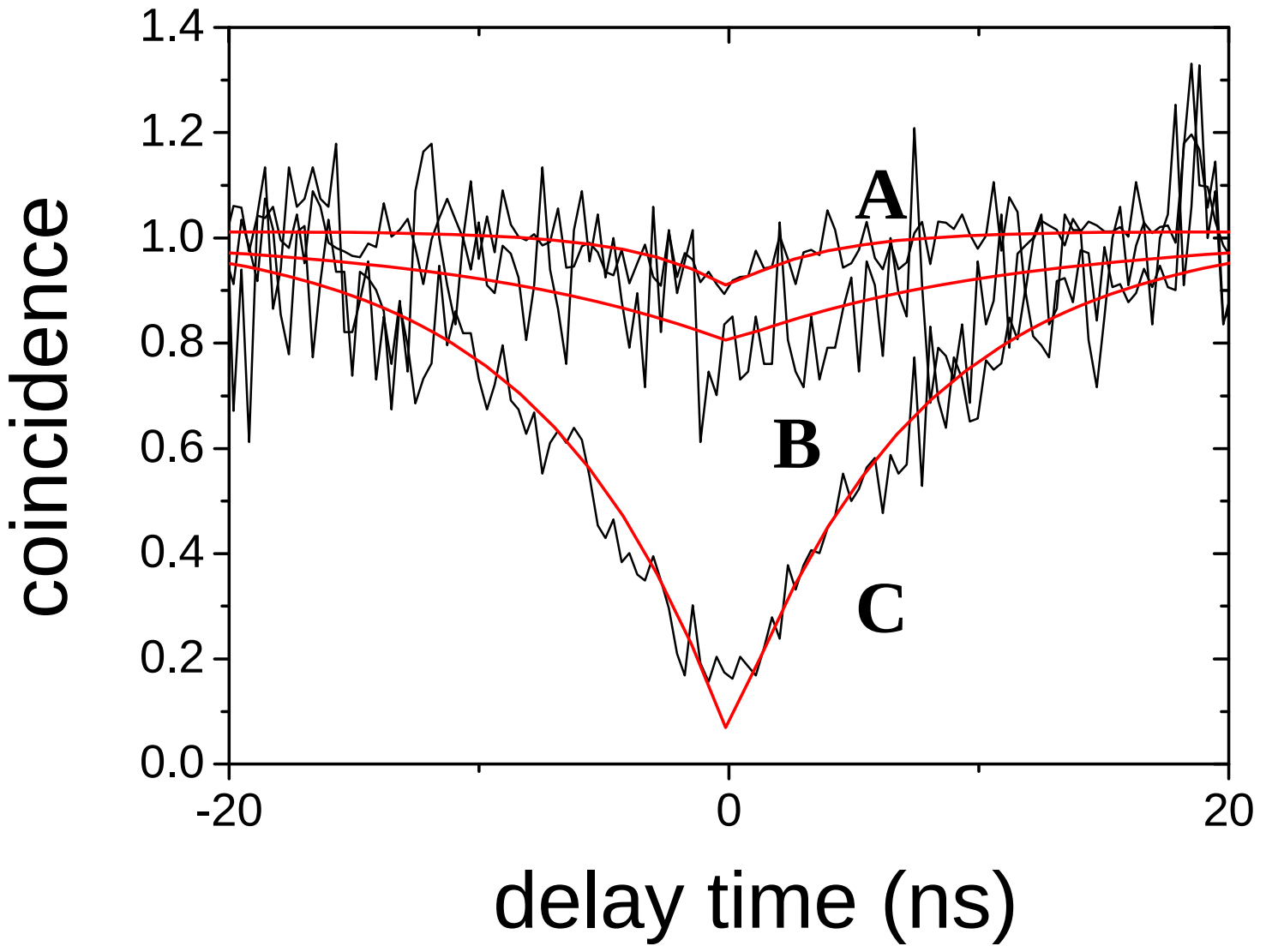

FIG. 2. J. R. Rabeau

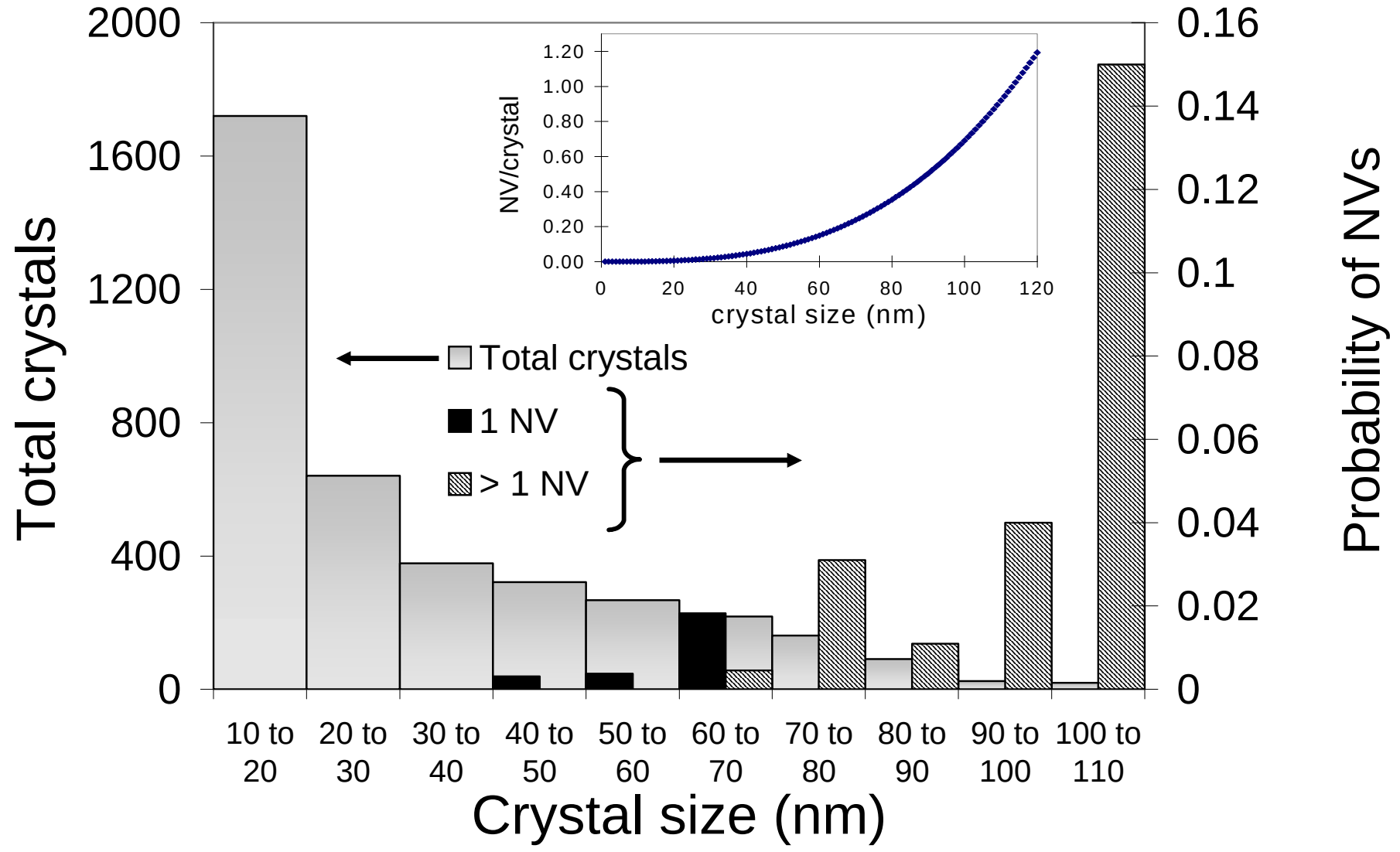

FIG. 3. J. R. Rabeau

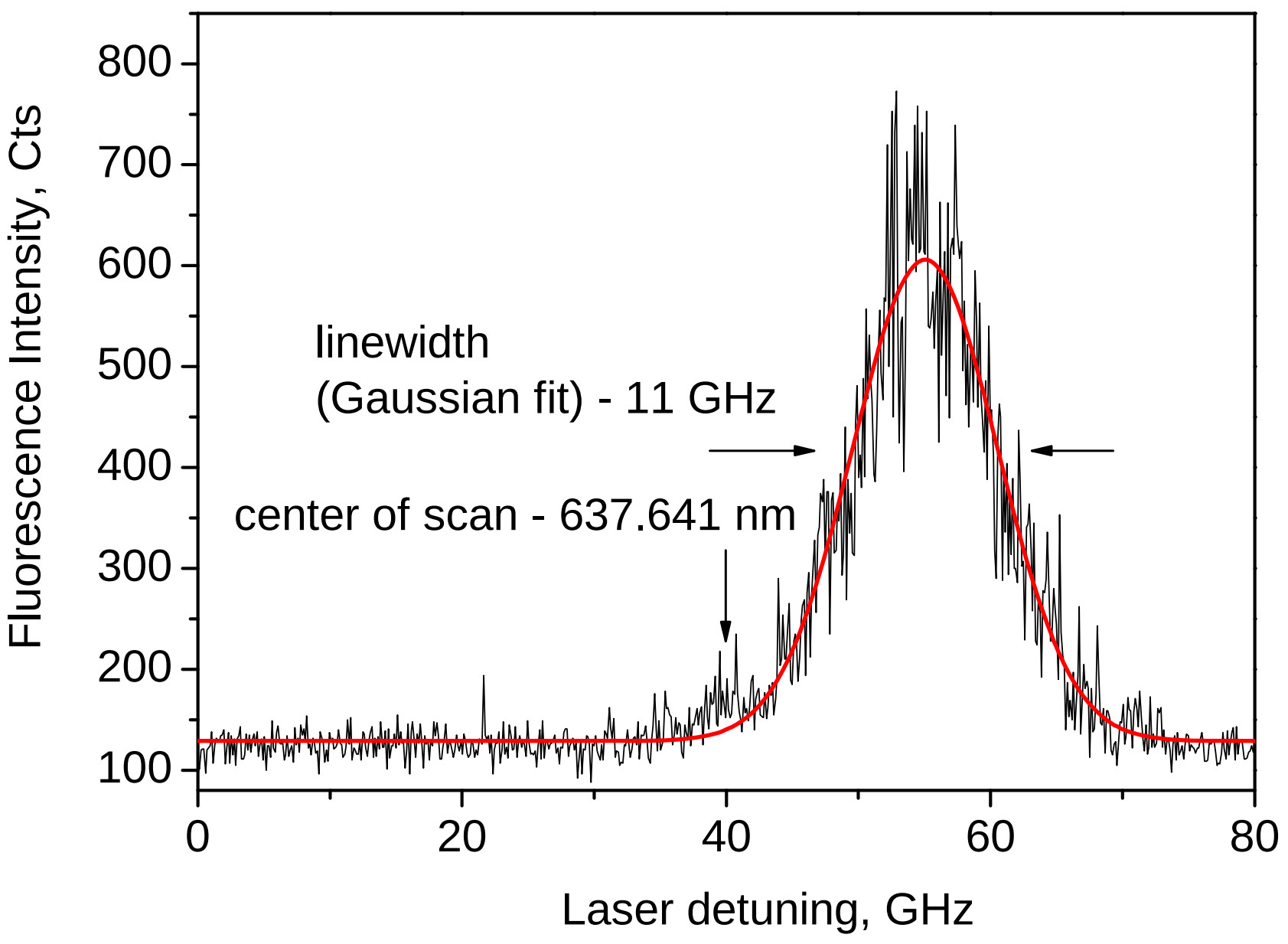

FIG 4. J. R. Rabeau

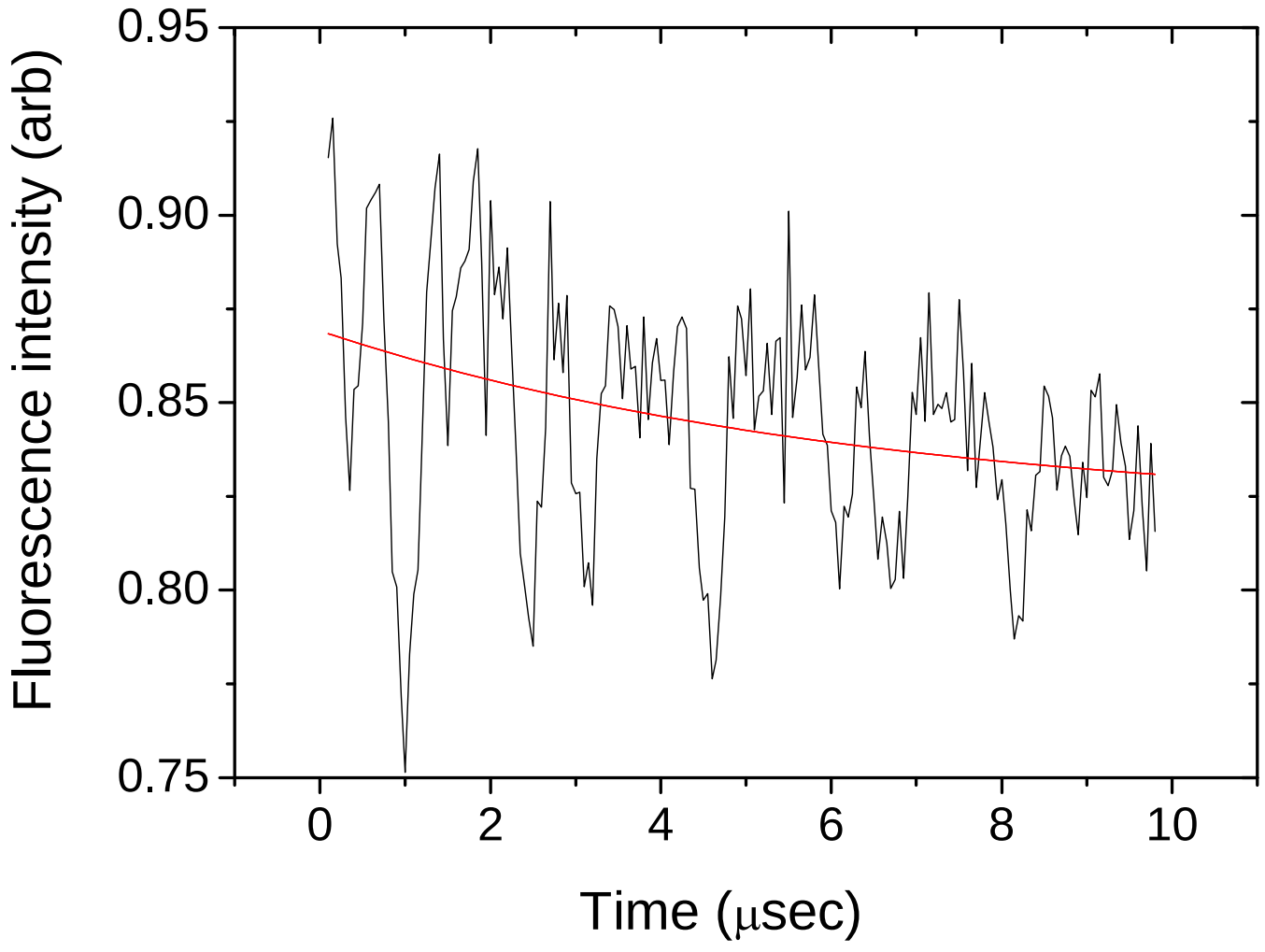

FIG 5. J. R. Rabeau